\documentclass[aps,prd,twocolumn,superscriptaddress,nofootinbib]{revtex4-2}

\usepackage{amsmath,amssymb}
\usepackage{graphicx}
\usepackage{booktabs,array,multirow}
\usepackage{subcaption}
\usepackage{xcolor}
\usepackage[utf8]{inputenc}
\usepackage{newunicodechar}
\newunicodechar{−}{-}
\newunicodechar{⊙}{\ensuremath{\odot}}
\usepackage{hyperref}

\IfFileExists{siunitx.sty}{\usepackage{siunitx}\newcolumntype{N}{S[table-format=3.6]}}{\newcolumntype{N}{r}}

\providecommand{\citep}{\cite}
\providecommand{\citet}{\cite}

\begin{document}

\title{Imprints of Higgs-portal fermionic dark matter on neutron-star tidal deformability and the mass--radius slope}

\author{Monmoy Molla}
\email{monmoymolla2016@gmail.com}
\affiliation{Department of Physics, Aliah University, New Town, Kolkata 700160, West Bengal, India}

\author{Mehedi Kalam}
\email{mehedikalam.phys@aliah.ac.in}
\affiliation{Department of Physics, Aliah University, New Town, Kolkata 700160, West Bengal, India}

\author{Tuhin Malik}
\email{tuhin.malik@gmail.com}
\affiliation{Departamento de F\'isica, Universidade de Coimbra, 3004-516 Coimbra, Portugal}

\date{\today}

\begin{abstract}
We investigate the structure of neutron stars (NSs) admixed with fermionic dark matter (DM) using three density-dependent relativistic mean-field (DDRMF) functionals (DDME, DDB, and GDFM) for $\beta$-equilibrated nucleonic matter. Modeling DM as the lightest neutralino interacting via Higgs exchange, we treat the DM Fermi momentum $k_F^{\rm DM}$ as a control parameter in the range $0.02$-$0.06$ GeV. We solve the coupled mean-field and Tolman-Oppenheimer-Volkoff equations to obtain the mass-radius relation, maximum mass $M_{\rm max}$, radial sound speed  profile $c_s^2$, and tidal deformability $\Lambda$. In all models, DM softens the equation of state, systematically reducing $M_{\rm max}$, the radius $R_{1.4}$ (at $1.4M_{\odot}$), and the tidal deformability $\Lambda_{1.4}$ (at $1.4M_{\odot}$) as $k_F^{\rm DM}$ increases. Consequently, the $2 M_\odot$ pulsar limit and NICER data place a model-dependent upper limit on the DM content, while the GW170817 tidal bound requires a minimal DM content for the stiffest functional. Using a recent Bayesian inference of the DDRMF equation of state as the nucleonic reference band, we evaluate if this DM imprint can be distinguished from nucleonic uncertainties using only measureable quantities. Analyzing the tidal deformability $\Lambda$ and mass-radius slope $dR/dM$ at fixed mass, we find that $\Lambda$ is a sharp discriminator: at $1.4 M_\odot$ and $k_F^{\rm DM}=0.06$ GeV, the DM-induced reduction of $\Lambda$ reaches $\simeq 8$ times the nucleonic $1\sigma$ width (model-independently $7.5$-$7.8\sigma$). The DM track leaves the nucleonic $1\sigma$ band for $k_F^{\rm DM}\gtrsim0.03$-$0.04$ GeV, whereas $dR/dM$ becomes diagnostic only for the heavier ($1.8$-$2.0 M_\odot$) branch.
\end{abstract}

\maketitle
\vspace{1cm}
\section{Introduction}\label{sec:1}
Neutron stars (NSs), owing to their extreme compactness, are among the most remarkable objects in the
Universe and serve as natural laboratories for matter at densities several times the nuclear saturation
density $n_0$. Despite their central role in astrophysics, the composition and the equation of state
(EoS) of their dense cores remain poorly constrained \citep{Glendenning:1997wn,Lattimer:2000nx}.
A complementary open question of modern physics is the nature of dark matter (DM), whose gravitational
influence is firmly established by galactic rotation curves and large-scale structure
\citep{Rubin:1978kmz,Bauer:2017qwy} but whose particle identity is unknown. Because their interiors reach
the highest densities accessible in the cosmos, NSs can capture and accumulate DM from the surrounding
halo \citep{Bertone:2007ae,Goldman:1989nd,Kouvaris:2010vv}, and the trapped component can in turn
modify the macroscopic structure of the star \citep{Panotopoulos:2017idn,Das:2018frc}. NSs
therefore offer a unique astrophysical probe of DM that is complementary to terrestrial direct-detection
experiments.

A variety of DM candidates --- bosonic and fermionic, self-interacting or not --- have been examined in
the NS context \citep{Panotopoulos:2017idn,Das:2018frc,Thakur:2024btu}. This has grown into an active,
multi-group effort spanning a wide range of DM models and stellar frameworks. Fermionic-DM stars and
two-fluid dark-compact-planet configurations have been studied extensively by Tolos, Schaffner-Bielich
and collaborators \citep{Narain:2006kx,Tolos:2015qra,Dengler:2021qcq,Barbat:2024yvi}; self-interacting
bosonic DM and its imprint on tidal deformability have been explored by Sagun, Ivanytskyi and
co-workers \citep{Giangrandi:2022wht,Ivanytskyi:2019wxd}; and Bayesian, data-driven constraints on the
DM content have been developed by Rutherford et al.\ \citep{Rutherford:2022xeb}. These works, using
either bosonic or fermionic DM and different coupling and halo/core geometries, consistently find that a
trapped dark component alters the mass, radius and tidal response of the star, motivating the systematic
approach we pursue here.
In the fermionic case the degeneracy pressure of the DM Fermi gas can by itself support a stable
configuration, and the resulting modification of the host star depends on the DM particle mass, its
abundance, and the way it couples to baryons. A particularly economical and widely studied realization
is the Higgs-portal scenario, in which a fermionic DM particle (the lightest neutralino) does not couple
to nucleons directly but only through the exchange of the Standard-Model Higgs boson
\citep{Das:2018frc,Panotopoulos:2017idn}. Previous studies established the qualitative trend that
a larger DM content softens the EoS and lowers the maximum mass, radius and tidal deformability
\citep{Thakur:2024btu,Molla:2025wij}, while the inferred DM fractions and particle masses span a
broad range depending on the assumed baryonic EoS and the observational data employed. These earlier works,
including our own companion study \citep{Molla:2025wij}, established this softening trend and derived
bounds on the DM content by comparing DM-admixed sequences to one or a few fixed nucleonic EoSs. The
present work departs from that approach in two respects: we quantify the DM imprint against a full
\emph{Bayesian} posterior of the nucleonic EoS rather than a fixed reference, and we ask specifically
whether that imprint on measurable quantities can be separated from the nucleonic-EoS uncertainty itself
--- a distinguishability question, rather than a bound on the DM fraction.

The baryonic sector itself, however, is uncertain. Within the relativistic mean-field (RMF) framework,
the nucleon--meson interaction can be made density dependent in several ways, each consistent with
nuclear data but extrapolating differently to the supranuclear densities relevant for NS cores. Here we
adopt three such density-dependent RMF (DDRMF) parametrizations that differ in their functional forms:
the rational-function couplings of Typel \& Wolter \citep{Typel:1999yq} (denoted DDME), the
exponential couplings of Malik et al.\ \citep{Malik:2022zol} (DDB), and the generalized
functional form of G\"ogelein et al.\ \citep{Gogelein:2007qa}, with the parameter set of Char et
al.\ \citep{Char:2023fue} (GDFM). Comparing these three frameworks allows the
robustness of any DM signature against the choice of nucleonic functional to be assessed.

To place the DM effect in the context of the present nucleonic uncertainty, we use the publicly released
Bayesian DDRMF EoS posterior of Cartaxo et al.\ (2026)\,\citep{Cartaxo:2025jpi}, obtained with the
\textsc{CompactObject} inference framework \citep{Huang:2024rfg} and conditioned on nuclear
saturation properties, chiral-EFT and perturbative-QCD limits, the GW170817 tidal deformability, and the
NICER mass--radius measurements. This posterior provides a quantitative ``nucleonic uncertainty band''
against which a candidate DM imprint can be judged.

The aim of this work is twofold. First, we perform a systematic study of the structural properties of
Higgs-portal fermionic-DM-admixed NSs in the three DDRMF frameworks, varying the DM Fermi momentum
$k_F^{\rm DM}$ over $0.02$--$0.06$~GeV and confronting the results with the two-solar-mass pulsar limit,
the NICER radii of PSR~J0030$+$0451, PSR~J0740$+$6620, PSR~J0437$-$4715 and PSR~J0614$-$3329
\citep{Riley:2019yda,Riley:2021pdl,Choudhury:2024xbk,Mauviard:2025dmd}, and the GW170817 tidal bound
\citep{LIGOScientific:2018cki}, thereby deriving model-dependent upper limits on the DM content. Second, and
as our principal new contribution, we ask whether the DM imprint on quantities evaluated at fixed
gravitational mass ($1.4$, $1.8$ and $2.0\,M_\odot$) can be distinguished from the nucleonic EoS
uncertainty encoded in the Cartaxo posterior. We find that the dimensionless tidal deformability
$\Lambda$ --- a direct gravitational-wave observable --- is the sharpest discriminator, with the
mass--radius slope $dR/dM$ providing complementary leverage on the heavy-mass branch and the central
sound speed $c_s^2$ exposing the underlying mechanism (though it is not itself directly measurable). We compute all stellar properties using the \textsc{CompactObject}
toolkit, extending its density-dependent RMF module with a self-consistent Higgs-portal DM sector.

The paper is organized as follows. Section~\ref{sec:2} presents the formalism: the pure nucleonic DDRMF
EoS, the Higgs-mediated nucleonic matter (NM)--DM interaction, the EoS of the DM-admixed star, the TOV equations and the
tidal deformability. Section~\ref{sec:3} reports the numerical results and the multi-messenger
constraints, and presents the distinguishability analysis. Section~\ref{sec:conclusions} summarizes our
conclusions.

\section{Formalism}\label{sec:2}
In this section, we present the relativistic mean-field (RMF) model, which is employed to construct the EoSs for DM admixed NSs.
\subsection{Pure nuclear matter EoS}
\label{sec:2.1}
Determining the nuclear EoS essentially reduces to theoretically modeling nuclear interactions. In a phenomenological description, the effective nucleon–nucleon interaction is treated within a relativistic mean-field (RMF) framework, using an effective Lagrangian that contains the fields for mesons and baryons. In this picture, the interaction between neutron and proton is mediated by exchange of various mesons. The interaction of scalar mesons (\(\sigma\) meson) with the baryons is responsible for the  generation of a  central force which is  attractive in nature. This attractive force influences the spin–orbit potential. On the other hand, the interaction of baryons with the vector meson (\(\omega\) meson) accounts for the short-range repulsive force. In order to implement isospin symmetry and (approximate) isospin independence of the nuclear force, as well as to distinguish protons from neutrons, the isovector ($\rho$) meson is introduced. The Lagrangian incorporating the field of nucleons (protons and neutrons) and the various mesons  (\(\sigma\), \(\omega\), and \(\rho\) mesons), together with their mutual interactions, is expressed as

\begin{equation}
\begin{aligned}
\mathcal{L}^{NM} = &\bar{\Psi} \bigl[ \gamma^{\mu}(i\partial_{\mu} - \Gamma_{\omega}A_{\mu}^{(\omega)} - \Gamma_{\rho}\tfrac{\tau}{2} \cdot A_{\mu}^{(\rho)}) - (m - \Gamma_{\sigma}\phi) \bigr]\Psi \\
&+ \frac{1}{2}\bigl( \partial_{\mu}\phi \partial^{\mu}\phi - m_{\sigma}^{2}\phi^{2} \bigr) \\
&- \frac{1}{4}F_{\mu\nu}^{(\omega)}F^{(\omega)\mu\nu} + \frac{1}{2} m_{\omega}^{2} A_{\mu}^{(\omega)} A^{(\omega)\mu} \\
&- \frac{1}{4}F_{\mu\nu}^{(\rho)}\cdot F^{(\rho)\mu\nu} + \frac{1}{2} m_{\rho}^{2} A_{\mu}^{(\rho)}\cdot A^{(\rho)\mu},
\end{aligned}
\label{eq:1}
\end{equation}

where \(\Psi\) denotes the baryon spinors and, in this study, represents  isospin doublet for the nucleon (proton and neutron) with bare mass \(m\). The symbols \(\tau\) and \(\gamma^{\mu}\)  stand for the Pauli and Dirac matrices, respectively. The field strength tensors  are denoted as \(F^{(\omega,\rho)\mu\nu} = \partial^{\mu}A^{(\omega,\rho)\nu} - \partial^{\nu}A^{(\omega,\rho)\mu}\). In the DDRMF model, the meson–nucleon coupling constants depend on the baryon density, a feature first introduced by Brockmann and Toki \citep{brockmann1992relativistic}, which reflects the modification of the nucleon–nucleon (NN) interaction by the surrounding nuclear medium. The quantities \(\Gamma_{\sigma}\), \(\Gamma_{\omega}\), and \(\Gamma_{\rho}\) denote the density-dependent coupling strengths between the nucleons and the \(\sigma\), \(\omega\), and \(\rho\) meson fields, respectively, where the corresponding meson masses are \(m_{\sigma}\), \(m_{\omega}\), and \(m_{\rho}\). Generally, nucleon-meson couplings are prescribed as density-dependent parameters that have form

\begin{equation}
\Gamma_{M}(\rho) = \Gamma_{M,0}\,h_{M}(x),\quad x = \rho/\rho_{0},
\label{eq:2}
\end{equation}

where  \(\rho\) represents the density of baryons, \(\Gamma_{M,0}\) denotes the couplings at saturation density \(\rho_{0}\), and \(M \in \{\sigma, \omega, \rho\}\). Several studies have employed different choices for the isoscalar and isovector coupling parameters. For the couplings of isoscalar mesons, the function \(h_{M}\), as introduced by Typel \& Wolter in their work \citep{Typel:1999yq} (referred to as the \textit{DDME} model), is expressed as

\begin{equation}
    h_{M}(x) =a_{M} \frac{1+b_M(x+d_M)^2}{1+c_M(x+d_M)^2}, \quad M=\sigma,\omega
\end{equation}

The proposed form of isovector coupling is described as, 
\begin{equation}
h_{\rho}(x) = \exp\bigl[ -a_{\rho}(x - 1) \bigr].
\end{equation}
The number of independent parameters in the isoscalar coupling $h_M$ can be reduced to three by applying the constraints $h_M(1) = 1$, $h_M'' (0) = 0$, and $h_{\sigma}'' (1) = h_{\omega}'' (1)$. The first two constraints yield
\begin{equation}
    a_{M}= \frac{1+c_M(1+d_M)^2}{1+b_M(1+d_M)^2}, \quad 3c_Md_M^2=1
\label{eq:5}
\end{equation}

In the work of Malik et al.\ \citep{Malik:2022zol} (\textit{DDB} model), the isoscalar coupling is proposed in the following form
\begin{equation}
  h_{M}(x) = \exp\bigl[ - (x^{a_{M}} - 1) \bigr],\quad M = \sigma,\omega
\label{eq:6}
\end{equation}

The above parametrization specified in Equation (\ref{eq:6}) adds only a single additional parameter for each coupling, analogous to the \(\rho\)-meson coupling. It was selected so that the coupling between $\sigma$ and $\omega$ mesons with nucleons could exhibit a density dependence consistent with Dirac-Brückner-Hartree-Fock calculations \citep{Typel:1999yq,TerHaar:1986xpv,Brockmann:1990cn} for \(\rho \gtrsim 0.04\ \mathrm{fm}^{-3}\). This density interval is suitable for modeling the EoS of the NS core.

The generalized density-dependent functional form, originally introduced by G\"ogelein et al.\ \citep{Gogelein:2007qa} and later adopted (with the parameter set used here) by Char et al.\ \citep{Char:2023fue} --- typically referred to as the \textit{GDFM} model --- reads
\begin{equation}
   \Gamma_{M}(\rho) = a_M+(b_M+d_M x^3)e^{-c_M x},\quad M = \sigma,\omega ,\rho
\end{equation}

In the following, we use the mean field approximation and assume that the system is composed of static, homogeneous matter in its ground state . In this mean field approximation, the fields of mesons are replaced by their expectation values, $\langle \sigma \rangle$ and $\langle A_{\mu}^{(\omega,\rho)} \rangle$. We ignore the quantum fluctuations. The source densities and current densities $\bar{\psi}(x)\psi(x)$ and $\bar{\psi}(x)\gamma^{\mu}\psi(x)$ for static, uniform matter are independent of $x$.Furthermore, the only nonvanishing components of the vector fields are the time-like part of the \(\omega\) field, \(\omega_{0}\), and the third component of isospin of the  field of isovector meson ($\rho$ meson), \(\rho_{3}^{0}\). In our approximation, the Euler–Lagrange equations for all fields thus become

\begin{equation}
m_{\sigma}^{2}\sigma = \Gamma_{\sigma}\bar{\psi}\psi,\
\label{eq:8}
\end{equation}
\begin{equation}
m_{\omega}^{2}\omega_{0} = \Gamma_{\omega}\bar{\psi}\gamma_{0}\psi,\\
\label{eq:9}
\end{equation}
\begin{equation}
m_{\rho}^{2}\rho_{3}^{0} = \frac{1}{2}\Gamma_{\rho}\bar{\psi}\gamma_{0}\tau_{3}\psi .
\label{eq:10}
\end{equation}

At zero temperature, the baryon number density \(\rho = \langle \bar{\psi}\gamma^{0}\psi \rangle\) and the scalar density \(\rho_{s} = \langle \bar{\psi}\psi \rangle\) are given by

\begin{equation}
\rho = \frac{\gamma}{2\pi^{2}} \sum_{B=n,p} \int_{0}^{k_{FB}} k^{2} dk,\\
\label{eq:11}
\end{equation}
\begin{equation}
  \rho_{s} = \frac{\gamma}{2\pi^{2}} \sum_{B=n,p} \int_{0}^{k_{FB}} \frac{m^{*} k^{2}}{\sqrt{m^{*2}+k^{2}}} dk, 
\label{eq:12}  
\end{equation}

where $\gamma$ is the spin degeneracy factor and \(k_{FB}\) is the nucleon's Fermi momentum. The nucleon's effective mass is \(m^{*} = m - \Gamma_{\sigma}\sigma\). The chemical potential of the nucleons is given by \(\mu_{B} = \nu_{B} + \Gamma_{\omega}\omega_{0} + \Gamma_{\rho}\tau_{3B}\rho_{3}^{0} + \Sigma^{\prime}\). Here, \(\tau_{3B}\) represents the isospin projection, and the rearrangement term \(\Sigma^{\prime}\) accounts for many-body effects in the nuclear interaction \citep{Typel:1999yq} and guarantees thermodynamic consistency. This term arises as a result of the coupling constants' density dependency and is written as

\begin{equation}
\Sigma^{\prime} = \sum_{B=n,p} \left[ -\frac{\partial\Gamma_{\sigma}}{\partial\rho_{B}}\sigma \rho_{sB} + \frac{\partial\Gamma_{\omega}}{\partial\rho_{B}}\omega_{0}\rho_{B} + \frac{\partial\Gamma_{\rho}}{\partial\rho_{B}}\rho_{3}^{0}\tau_{3B}\rho_{B} \right],
\label{eq:13}
\end{equation}

where $\rho_{sB}$ is the scalar density of baryon species $B$ (so that $\rho_s=\sum_B\rho_{sB}$), $\tau_{3B}$ is its isospin projection, and the derivatives $\partial\Gamma_M/\partial\rho_B$ are evaluated at the total baryon density; the $\sigma$ contribution enters through the scalar density $\rho_{sB}$ rather than the vector density $\rho_B$, consistent with the definition of $\Sigma^{\prime}$ in Ref.~\citep{Typel:1999yq}.

The NM energy density is defined as

\begin{equation}
\begin{aligned}
\epsilon_{NM} = &\frac{1}{\pi^{2}} \sum_{B=n,p} \int_{0}^{k_{FB}} k^{2}\sqrt{k^{2}+m^{*2}} dk + \frac{1}{2} m_{\sigma}^{2}\sigma^{2} \\
&+ \frac{1}{2} m_{\omega}^{2}\omega_{0}^{2} + \frac{1}{2} m_{\rho}^{2}(\rho_{3}^{0})^{2} + \epsilon_{\mathrm{lep}},
\end{aligned}
\label{eq:14}
\end{equation}

where $\epsilon_{lep}$ represents the contribution from leptons(i.e., from electrons and muons) and is expressed as
\begin{equation}
  \epsilon_{lep} = \sum_{l=e,\mu} \frac{1}{\pi^2} \int_0^{k_l} k^2 \sqrt{k^2 + m_l^2} \, dk 
\label{eq:15}
\end{equation}

The Euler relation is used to determine the pressure \(P\) of NM from the energy density,

\begin{equation}
P_{NM} = \sum_{i=n,p,e,\mu} \mu_{i}\rho_{i} - \epsilon,
\label{eq:16}
\end{equation}

where \(\mu_{i}\) and \(\rho_{i}\) represent the chemical potential and the particle number density of species \(i\), respectively.

At its center, the star consists predominantly of neutrons occupying extremely high-momentum states. Beta decay (\(\beta\)-decay) maintains an equilibrium among protons, neutrons, muons, and electrons.

\begin{equation}
\begin{aligned}
n &\leftrightarrow p + e^{-} + \bar{\nu},\\
\end{aligned}
\label{eq:17}
\end{equation}
\begin{equation}
\begin{aligned}
n + \nu &\leftrightarrow p + e^{-},
\end{aligned}
\label{eq:18}
\end{equation}

and muons (\(\mu\)) are produced once the electron chemical potential reaches rest mass value of the muon (\(m_{\mu}=106\ \mathrm{MeV}\)). Neutrinos can effortlessly escape from a cold, catalytic NS because their wavelengths are substantially longer than the star radius. Consequently, the condition of \(\beta\)-equilibrium  takes the form

\begin{equation}
\mu_{n} = \mu_{p} + \mu_{e}  \hspace{1.9mm}and \qquad \mu_{\mu} = \mu_{e}.
\label{eq:19}
\end{equation}

The  charge neutrality condition considering  a constant baryon density ($\rho = \rho_{n}+\rho_{p}$) is given by,

\begin{equation}
 \rho_{p} = \rho_{e} + \rho_{\mu}.
 \label{eq:20}
\end{equation}

To determine properties of NSs, the crust EoS must be matched to the EoS for the core. We use the Bethe-Pethick-Sutherland (BPS) EoS for the outer crust. The polytropic relation \(P(\varepsilon) = a_{1} + a_{2}\varepsilon^{\gamma}\)\citep{Carriere:2002bx} connects the outer crust and the core. 
It is possible to determine the unknown parameters ($a_1$ and $a_2$) so that the edge of the core at one end (at $\rho\,=\,0.04\,fm^{-3}$) matches with the inner portion of the outer crust (at $\rho\,=\,10^{-4}\,fm^{-3}$) at the other.  
We set the polytropic index  \(\gamma\) equal to \(4/3\).

\subsection{The interaction Lagrangian between NM and DM}
\label{sec:2.2}
Owing to galactic rotation, compact objects such as NSs move through the surrounding DM halo and can capture DM particles from it. Inside a NS, the high baryon density causes DM particles to lose energy through interactions with neutrons. Once a DM particle has lost sufficient energy, the strong gravitational field of the NS gravitationally binds it \citep{Kouvaris:2007ay,Xiang:2013xwa,Goldman:1989nd}. The DM density within NSs can be further increased by other processes, such as the generation of scalar DM by bremsstrahlung and the conversion of neutrons into scalar DM \citep{Ellis:2018bkr,Ellis:2017jgp}. Given that DM constitutes about \(95\%\) of the total density of matter, it is conceivable that many compact objects contain a substantial DM component. The overall quantity of DM inside a NS also depends on its evolutionary history and the properties of its environment. In Ref. \citep{Brayeur:2011yw}, the authors demonstrate that binary NS systems can enhance the probability of DM accumulation within NSs.

We focus on fermionic DM \((\chi)\) inside a NS, taking the lightest neutralino as the representative fermionic DM candidate \citep{Panotopoulos:2017idn,Murakami:2000me}. The DM does not couple directly to nucleons; instead, it interacts via the Higgs field \(h\). The corresponding DM–Higgs coupling is denoted by \(y\). For a neutralino with mass \(m_{\chi} = 200\ \mathrm{GeV}\), the parameter \(y\) is constrained to the interval 0.001–0.1, and in the rest of our analysis we take \(y = 0.07\) as a benchmark value \citep{Panotopoulos:2017idn,Murakami:2000me}. The Higgs field also interacts with nucleons via an effective Yukawa term \(\frac{f m}{v}\), where \(f\) denotes the proton–Higgs form factor, which is estimated to be about 0.35 \citep{Cline:2013gha}. Since the expectation value of $h$ is negligibly small under the mean-field approximation, we ignore the contributions of $h^3$ and $h^4$ in the Higgs potential, leaving the quadratic $h^2$ term as the dominating contribution. Consequently, the DM Lagrangian, along with its couplings to nucleons and the Higgs field, can be expressed as \citep{Panotopoulos:2017idn}

\begin{equation}
\begin{array}{l}
{\mathcal{L}_{\mathrm{DM}}=\bar{\chi}[i\gamma^{\mu}\partial_{\mu}-m_{\chi}+y h]\chi+\frac{1}{2}\partial_{\mu}h\partial^{\mu}h-\frac{1}{2}m_{h}^{2}h^{2}} \\
{{}+f\frac{m_{}}{v}\bar{\psi}h\psi.}
\end{array} 
\label{eq:21}
\end{equation}

For DM masses in the $30$–$50$ GeV range at $90\%$ confidence level, the current generation of direct-detection experiments, XENONnT \citep{XENON:2023cxc} and LUX-ZEPLIN (LZ) \citep{LZ:2022lsv}, have ruled out spin-independent DM–nucleon cross sections greater than \(\sim 10^{-47}\ \mathrm{cm}^2\) (with a minimum sensitivity of \(\sim 9\times 10^{-48}\ \mathrm{cm}^2\) near $30$–$40$~GeV). The constraint from the invisible Higgs decay width is very restrictive for DM masses below \(m_{h}/2\) \citep{Arcadi:2019lka}; therefore, a DM mass of \(m_{\chi} = 200\) GeV is not affected by these bounds. It should also be emphasized that DM need not be composed of a single species; it can naturally be a multicomponent system. In particular, DM may contain both light and heavy particles. For instance, in Ref. \citep{Panotopoulos:2017idn} the authors consider heavy DM particles residing inside NSs.

Here, we assume that the average number density of NM is \(10^{3}\) times greater than the average DM number density \((n_{\mathrm{DM}})\). Under this assumption, the DM mass fraction, which is basically ratio of the mass of DM inside NS to the total NS mass is \(\sim \frac{1}{6}\) \citep{Panotopoulos:2017idn}. Since the nuclear saturation density is \(n_{B} \sim 0.16\ \mathrm{fm}^{-3}\), the corresponding DM number density becomes \(n_{\mathrm{DM}} \sim 10^{-3} n_{B} \sim 0.16 \times 10^{-3}\ \mathrm{fm}^{-3}\). Using the relation \(n_{\mathrm{DM}} = \frac{(k_{\mathrm{F}}^{\mathrm{DM}})^{3}}{3\pi^{2}}\), we obtain \(k_{\mathrm{F}}^{\mathrm{DM}} \sim 0.033\ \mathrm{GeV}\). This capture-based estimate motivates our choice to treat \(k_{\mathrm{F}}^{\mathrm{DM}}\) as a control parameter over the range \(0.02\)–\(0.06\)~GeV, which brackets the physically expected value while spanning DM number densities \(n_{\mathrm{DM}}\sim(0.2\)–\(6)\times10^{-3}\,n_B\); rather than committing to a single accreted abundance --- which depends on the poorly known capture history and environment --- we map out the sensitivity of the observables across this range, so that the derived bounds are stated as functions of \(k_{\mathrm{F}}^{\mathrm{DM}}\) (equivalently, of the DM number density) and not of a single assumed relic fraction.

The DM Lagrangian density provided in Eq.(\ref{eq:21}) can be used to derive the Euler–Lagrange equation of motion for the DM particle $\chi$ and the Higgs boson $h$. Under the mean-field approximation, this equation takes the form
\begin{equation}
   h_0 = \frac{y\langle\bar{\chi}\chi\rangle + f\frac{m}{v}\langle\bar{\psi}\psi\rangle}{m_h^2},
\label{eq:22}
\end{equation}

The masses of the DM particle and the Higgs particle are represented by $m_{\chi}$ and $m_h$, respectively. The  DM scalar density \(\rho_s^{\mathrm{DM}}\) is given as,

\begin{equation}
\rho_s^{\mathrm{DM}} = \langle \bar{\chi}\chi \rangle = \frac{\gamma}{(2\pi)^3}\int_0^{k_F^{\mathrm{DM}}}\frac{m_{\chi}^*}{\sqrt{m_\chi^{*2} + k^2}}\, d^3 k,
\label{eq:23}
\end{equation}

where, \(k_{F}^{\mathrm{DM}}\) denotes the Fermi momentum of the DM 
\begin{table}[htbp]
\centering
\caption{Masses of nucleons and mesons, meson coupling strengths, and nuclear saturation densities used in different DDRMF models.}
\begin{tabular}{l N N N}
\toprule
& \multicolumn{1}{c}{DDME\,\citep{Typel:1999yq}} & \multicolumn{1}{c}{DDB \,\citep{Malik:2022zol}} & \multicolumn{1}{c}{GDFM\,\citep{Char:2023fue}} \\
\midrule
$m_n$ [MeV]      & 939.565  & 939.565  & 939.000  \\
$m_p$ [MeV]      & 938.272  & 938.272  & 939.000  \\
$m_\sigma$ [MeV] & 547.333 & 550.000 & 550.000  \\
$m_\omega$ [MeV] & 783.000   & 783.000   & 782.600   \\
$m_\rho$ [MeV]   & 763.000   & 763.000   & 769.000   \\
\midrule
$\Gamma_\sigma(0)$  & 10.706 & 9.180 & {\textendash}   \\
$\Gamma_\omega(0)$  & 13.338 & 10.981 & {\textendash}   \\
$\Gamma_\rho(0)$    & 7.238 & 7.653   &   {\textendash}  \\
\midrule
$\rho_{B0}$ [fm$^{-3}$] & 0.149 & 0.150    & 0.1619    \\
\midrule
$a_\sigma$ & 1.3970 & 0.0864 & 8.2255   \\
$b_\sigma$ & 1.3350 & {\textendash} & 2.7079   \\
$c_\sigma$ & 2.0671 & {\textendash} & 2.4776   \\
$d_\sigma$ & 0.4016 & {\textendash} & 3.8630   \\
$a_\omega$ & 1.3936 & 0.0541 & 10.4267   \\
$b_\omega$ & 1.0191 & {\textendash} & 1.6468   \\
$c_\omega$ & 1.6060 & {\textendash} & 6.8349   \\
$d_\omega$ & 0.4556 & {\textendash} & 1.4458  \\
$a_\rho$ & 0.6202 & 0.5092 & 0.6458   \\
$b_\rho$ & {\textendash} & {\textendash} & 5.2033   \\
$c_\rho$ & {\textendash} & {\textendash} & 0.4262   \\
$d_\rho$ & {\textendash} & {\textendash} &  -0.1824   \\

\bottomrule
\end{tabular}
\label{tab:ddrmf}
\end{table}
particles. The symbol \(\gamma\) represents the spin degeneracy factor of the nucleons, with \(\gamma = 2\) for both neutrons and protons separately. \(m_{\chi}^{*}\) denotes the effective mass of the DM and can be expressed as
\begin{equation}
    m_\chi^* = m_\chi - y h_0,
\label{eq:24}
\end{equation}
It should be noted that the effective nucleon mass, $m^*$, deviates slightly from the expression given in Sec. \ref{sec:2.1} because of the interaction between NM and DM. In the presence of NM–DM interaction, it can be written as
\begin{equation}
   m^* = m - \Gamma_{\sigma}\sigma - \frac{fm}{v} h_0 
\label{eq:25}
\end{equation}
\subsection{Equation of state for DM admixed NS}
\label{sec:2.3}
For convenience, we now rewrite all the equations of motion (for both NM and DM) in the mean-field approximation, as obtained in Sec. \ref{sec:2.1} and Sec. \ref{sec:2.2}.

\begin{equation*}
 m_{\sigma}^{2}\sigma = \Gamma_{\sigma}\bar{\psi}\psi,\\
 \end{equation*}
 \begin{equation*}
 m_{\omega}^{2}\omega_{0} = \Gamma_{\omega}\bar{\psi}\gamma_{0}\psi,\\
\end{equation*}
\begin{equation*}
  m_{\rho}^{2}\rho_{3}^{0} = \frac{1}{2}\Gamma_{\rho}\bar{\psi}\gamma_{0}\tau_{3}\psi  
\end{equation*}
\begin{equation}
    h_0 = \frac{y\langle\bar{\chi}\chi\rangle + f\frac{m}{v}\langle\bar{\psi}\psi\rangle}{m_h^2},
    \label{eq:26}
\end{equation}
where the notation has already been introduced in the previous subsection. The baryon density $\rho$ and the nucleonic scalar density $\rho_s$ retain the forms given in Eqs.\ (\ref{eq:11}) and (\ref{eq:12}), while the DM scalar density is
\begin{equation}
\rho_s^{DM}  = \langle \bar{\chi}\chi \rangle = \frac{\gamma}{(2\pi)^3}\int_0^{k_F^{\mathrm{DM}}}\frac{m_{\chi}^*}{\sqrt{m_\chi^{*2} + k^2}}\, d^3 k .
\label{eq:27}
\end{equation}

For fixed values of the DM Fermi momentum \(k_{F}^{\mathrm{DM}}\) and the coupling constants, the masses of both the nucleon and DM depend on the baryon density. The corresponding masses and coupling constants are summarized in Table \ref{tab:ddrmf} and discussed in Sec.\ref{sec:2.2}. The density dependence of \(m^{*}\) and \(m_{\chi}^{*}\) is obtained by numerically solving Eq. (\ref{eq:27}) as well as the field equations, Eq. (\ref{eq:26}), in a self-consistent way. The average DM number density in this study is assumed to be approximately 1000 times lower than the average neutron number density. This choice leads to a DM mass fraction of \(\simeq 1/6\) relative to the total NS mass. In the static case, the expectation values of the energy‑momentum (stress) tensor yield the system’s energy density and pressure, i.e., the EoS, which is given by,

\begin{equation}
\epsilon = \langle T^{00}\rangle \quad \text{and} \quad P = \frac{1}{3}\langle T^{ii}\rangle.
\label{eq:28}
\end{equation}

The formulas for the total energy density \(\epsilon\) and the total pressure \(P\) for the DM-admixed NS are derived by combining the Lagrangian densities given in Eqs.\ (\ref{eq:1}) and (\ref{eq:21}):

\begin{equation}
\begin{aligned}
\epsilon_{} &= \frac{1}{\pi^{2}} \sum_{B=n,p} \int_{0}^{k_{FB}} k^{2}\sqrt{k^{2}+m^{*2}} dk \\
&\quad + \frac{1}{\pi^{2}}\int_{0}^{k_{F}^{\mathrm{DM}}} dk\, k^{2}\sqrt{k^{2}+(m_{\chi}^{*})^{2}} \\
&\quad + \sum_{l=e,\mu} \frac{1}{\pi^2} \int_0^{k_l} k^2 \sqrt{k^2 + m_l^2} \, dk \\
&\quad + \frac{1}{2}m_{\sigma}^{2}\sigma^{2} + \frac{1}{2}m_{\omega}^{2}\omega_{0}^{2} + \frac{1}{2}m_{\rho}^{2}(\rho^0_{3})^{2} + \frac{1}{2}m_{h}^{2}h_{0}^{2},
\end{aligned}
\label{eq:29}
\end{equation}

\begin{equation}
\begin{aligned}
P_{} &= \frac{1}{3\pi^{2}} \sum_{B=n,p} \int_{0}^{k_{FB}} \frac{k^{4}}{\sqrt{k^{2}+m^{*2}}} dk \\
&\quad +\frac{1}{3\pi^{2}} \sum_{l=e,\mu} \int_{0}^{k_{l}} \frac{k^{4}}{\sqrt{k^{2}+m_{l}^{2}}} dk \\
&\quad + \frac{1}{3\pi^{2}}\int_{0}^{k_{F}^{\mathrm{DM}}} \frac {k^{4}}{\sqrt{k^{2}+(m_{\chi}^{*})^{2}}}\,dk  \\
&\quad - \frac{1}{2}m_{\sigma}^{2}\sigma^{2} + \frac{1}{2}m_{\omega}^{2}\omega_{0}^{2} + \frac{1}{2}m_{\rho}^{2}(\rho^0_{3})^{2} - \frac{1}{2}m_{h}^{2}h_{0}^{2}\\
&\quad + \rho\,\Sigma^{\prime},
\end{aligned}
\label{eq:30}
\end{equation}

\subsection{Hydrostatic configuration}
\label{sec:2.4}

The metric describing a static, spherically symmetric spacetime takes the form \citep{Tolman:1939jz,Oppenheimer:1939ne}
\begin{equation}
ds^2 = -e^{\nu(r)}dt^2 + e^{\lambda(r)}dr^2 + r^2(d\theta^2 + \sin^2\theta d\varphi^2).
\label{eq:31}
\end{equation}
The structure of a nonrotating compact star in equilibrium is determined by the Tolman–Oppenheimer–Volkoff (TOV) equation, obtained by solving the Einstein field equations for the metric specified above and the conservation equation of energy-momentum tensor. The TOV equations, expressed in natural units where $G = c = 1$, take the form \citep{Oppenheimer:1939ne}:
\begin{equation}
\frac{dP}{dr} = -\frac{m(r) + 4\pi r^3 p(r)}{r^2(1 - 2m(r)/r)} (\epsilon(r) + P(r)), 
\label{eq:32}
\end{equation}
\begin{equation}
\frac{dm}{dr} = 4\pi r^2 \epsilon(r), 
\label{eq:33}
\end{equation}
\begin{equation}
\frac{d\nu}{dr} = \frac{2(m(r) + 4\pi r^3 p(r))}{r^2(1 - 2m(r)/r)}, 
\label{eq:34}
\end{equation}
where \(\epsilon\) and \(P\) are the energy density and pressure, respectively. The mass function \(m(r)\) is introduced via the relation \(e^{-\lambda(r)} = 1 - 2m(r)/r\). To obtain a solution of the TOV equation, we must specify suitable boundary conditions and a corresponding EoS, \(P(\epsilon)\). At the stellar center, the boundary conditions are \(m(r=0) = 0\) and \(\epsilon(r=0) = \epsilon_c\), where \(\epsilon_c\) denotes the chosen central energy density. Next, the TOV equation is integrated outward from $r = 0$ to $r = R$, where $R$ is specified as the star's radius by the constraint $P(R) = 0$.  We obtain \(\frac{dP}{dr} \to 0\) as \(r \to 0\) by taking the proper limit of the right-hand side of Eq. (\ref{eq:32}). The metric function \(\nu(r)\) satisfies the surface boundary condition \(e^{\nu(R)} = 1 - 2M/R\).

\subsection{Tidal deformability and tidal Love number}
A key structural property of a binary merger that can be quantified is its tidal deformability. In a merging binary NS system, a mass quadrupole moment is induced in the NS during the late inspiral stage. This occurs due to the tidal gravitational field produced by its companion NS in the binary. The tidal deformability measures how much an NS is deformed by the tidal field created by its companion NS, and is very sensitive to the properties of the EoS. The tidal deformability is then  defined as,
\begin{equation}
\lambda = \frac{2}{3}k_2R^5, 
\label{eq:35}
\end{equation}
where \( R \) is the radius of the NS. For NSs, the value of \( k_2 \) usually lies in the interval \(\simeq 0.05-0.15\) \citep{Hinderer:2007mb,PhysRevD.81.123016,Postnikov:2010yn} and is determined by the internal stellar structure. It can be evaluated from the following expression \citep{Hinderer:2007mb}
\begin{equation}
\begin{split}
k_2 =& \frac{8C^5}{5}(1 - 2C)^2[2 + 2C(y_R - 1) - y_R] \\
&\times \bigl\{ 2C(6 - 3y_R + 3C(5y_R - 8)) \\
&\quad + 4C^3[13 - 11y_R + C(3y_R - 2) + 2C^2(1 + y_R)] \\
&\quad + 3(1 - 2C)^2[2 - y_R + 2C(y_R - 1)] \log(1 - 2C) \bigr\}^{-1},
\end{split}
\label{eq:36}
\end{equation}
where $C\equiv M/R$ is the compactness parameter of a star with mass $M$.  The value \( y_R \equiv y(R) \) is determined by solving the following differential equation
\begin{equation}
r \frac{dy(r)}{dr} + y(r)^2 + y(r)F(r) + r^2Q(r) = 0 
\label{eq:37}
\end{equation}
with
\begin{equation}
F(r) = \frac{r - 4\pi r^3(\epsilon(r) - P(r))}{r - 2m(r)},
\label{eq:38}
\end{equation}
\begin{equation}
\begin{split}
Q(r) =& \frac{4\pi r\left(5\epsilon(r) + 9P(r) + \dfrac{\epsilon(r) + P(r)}{\partial P(r)/\partial \epsilon(r)} - \dfrac{6}{4\pi r^2}\right)}{r - 2m(r)} \\
&- 4\left[\frac{m(r) + 4\pi r^3P(r)}{r^2(1 - 2m(r)/r)}\right]^2,
\label{eq:39}
\end{split}
\end{equation}
together with the TOV equation supplemented by appropriate boundary conditions \citep{Oppenheimer:1939ne,Malik:2018zcf}. This allows one to introduce the dimensionless tidal deformability,
\( \Lambda = \frac{2}{3}k_2C^{-5} \).

\section{Results}
\label{sec:3}
\begin{figure*}[!th]
    \centering
    \begin{tabular}{cc}
       \includegraphics[width=0.95\linewidth]{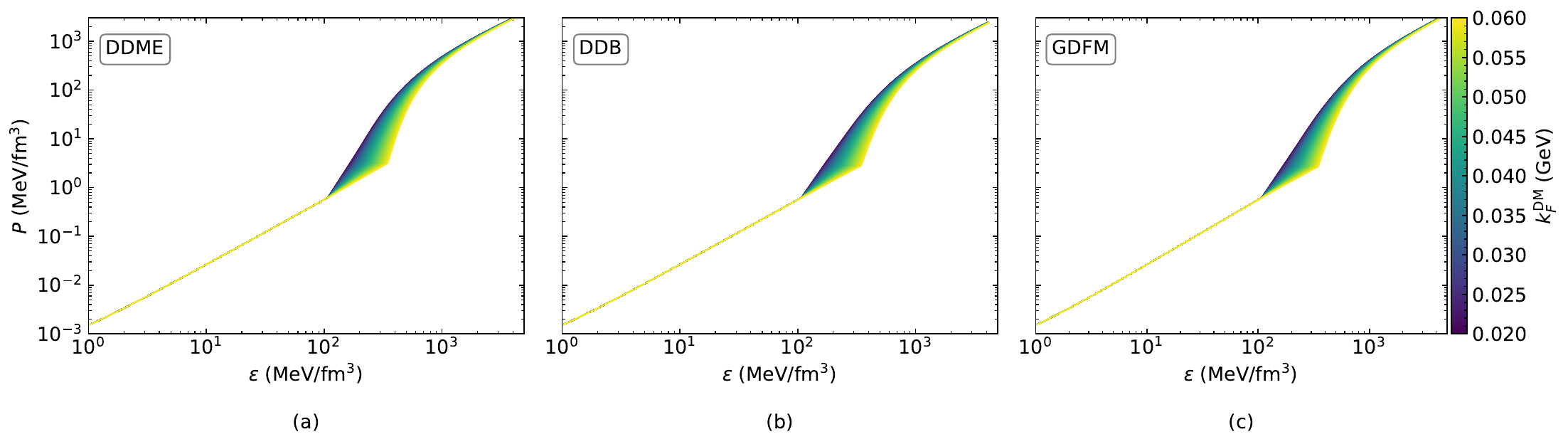}    
    \end{tabular}
    \caption{The EoSs of NSs for various DM Fermi momenta $k_F^{DM}$, ranging from 0.02 to 0.06 GeV, are shown for three density-dependent RMF models. As the DM number density inside the NS increases, the EoS becomes softer. Both axes are logarithmic.}
    \label{fig:1}
\end{figure*}

In  Fig. \ref{fig:1}, we show the pressure (P) as a function of the total energy density ($\epsilon$) for several values of the DM Fermi momentum, $k_F^{DM}=0.02$–$0.06$ GeV, for the three different models. For all three models, increasing $k_F^{DM}$ from 0.02 GeV to 0.06 GeV softens the EoS; that is, as the DM density rises, the pressure decreases, in agreement with previous studies \citep{Panotopoulos:2017idn,daSilva:2017swg}. This can be seen from the expressions for the energy density and pressure in Eqs. (\ref{eq:29}) and (\ref{eq:30}), in particular, as $k_F^{DM}$ increases the DM contribution to the energy density increases much faster than its contribution to the pressure.

\begin{figure*}[!th]
    \centering
    \begin{tabular}{cc}
       \includegraphics[width=0.95\linewidth]{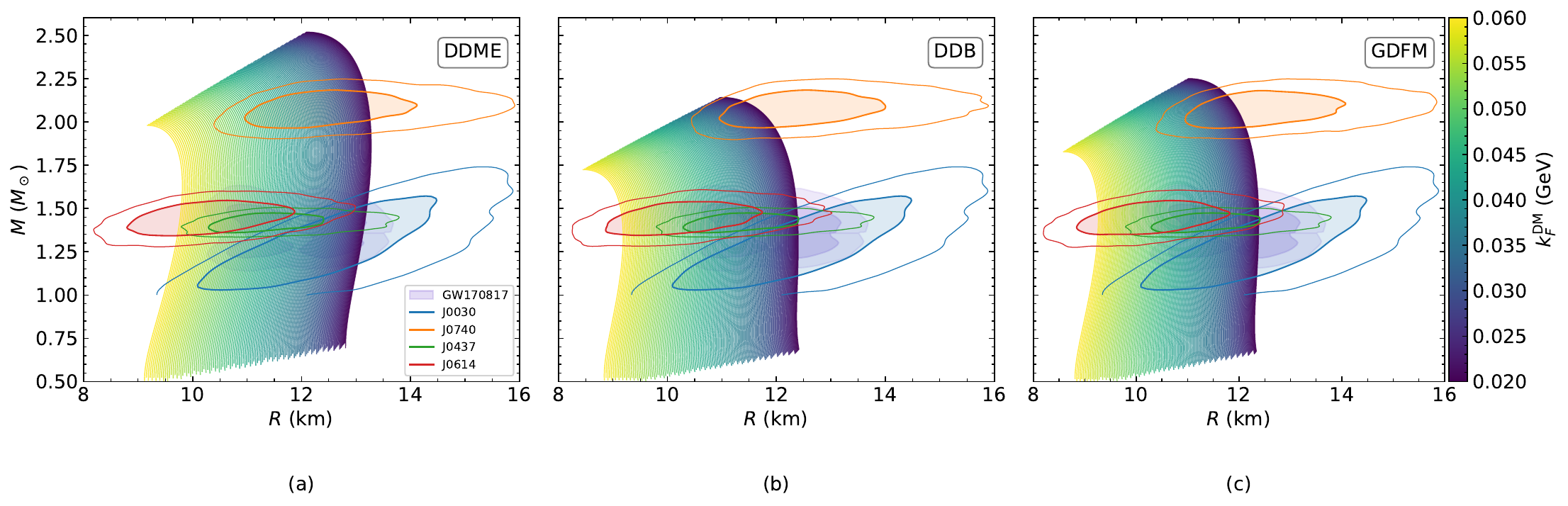}    
    \end{tabular}
    \caption{Mass–radius curves of NSs are shown for three different EoS models, assuming a DM Fermi momentum of $k_F^{\rm DM}=0.02$–$0.06$ GeV. The filled regions show the NICER credible regions (68\%, with 95\% outlines) for PSR J0030+0451 (blue) \citep{Riley:2019yda}, PSR J0740+6620 (orange) \citep{Riley:2021pdl}, PSR J0437−4715 (green) \citep{Choudhury:2024xbk}, and PSR J0614−3329 (red) \citep{Mauviard:2025dmd}. The purple shaded region shows the 50\% and 90\% credible regions of GW170817 \citep{LIGOScientific:2018cki}.}
    \label{fig:2}
\end{figure*}

The mass–radius relation of NSs is determined by solving the TOV equation for the EoSs described above. In Fig.~\ref{fig:2}, we display the NS mass–radius relations corresponding to the three EoS models introduced in Fig.~\ref{fig:1} for DM Fermi momenta ranging from $k_F^{DM}=0.02$ GeV to $0.06$ GeV . The colored contours indicate the allowed ranges of mass and radius inferred from astronomical observations. The blue, orange, green, and red filled regions represent the $1\sigma$ confidence intervals for the pulsars \textit{PSR J0030+0451} \citep{Riley:2019yda}, \textit{PSR J0740+6620} \citep{Riley:2021pdl}, \textit{PSR J0437-4715} \citep{Choudhury:2024xbk}, and \textit{PSR J0614-3329} \citep{Mauviard:2025dmd}, respectively. The purple shaded region shows the 50\% and 90\% credible regions of GW170817 \citep{LIGOScientific:2018cki}. In all three models, a clear and uniform trend is observed, in which the maximum mass and its associated radius decrease with increasing $k_F^{DM}$, thus directly reflecting the EoS softening shown in Fig.\ref{fig:1}. For $k_F^{DM}=0.02$ GeV, all three models predict maximum stellar masses well above $2M_{\odot}$. Among them, DDME provides the heaviest stars with the largest radii, whereas DDB and GDFM give slightly lower, yet broadly similar, mass and radius values. However, once $k_F^{DM}$ nears $0.06$ GeV, the maximum mass predicted by all models decreases sharply, with DDB and GDFM dropping below the $2M_{\odot}$ limit—placing these results in conflict with the well established pulsar mass observations. Regarding consistency with observations, the low-DM-density curves for all three models largely coincide with the NICER posterior contours for PSR J0030+0451, PSR J0740+6620, PSR J0437–4715, and PSR J0614–3329, as well as the $1\sigma$  constraint from GW170817, suggesting that small but finite DM density remain  well consistent with observations. Conversely, configurations with large $k_F^{DM}$ yield radii that are overly compact and masses that are too small to satisfy the complete set of multi-messenger constraints at the same time, thereby imposing an effective upper limit on the allowed DM content in NSs within this fermionic DM scenario.

\begin{figure*}[!th]
    \centering
    \begin{tabular}{cc}
       \includegraphics[width=0.98\linewidth]{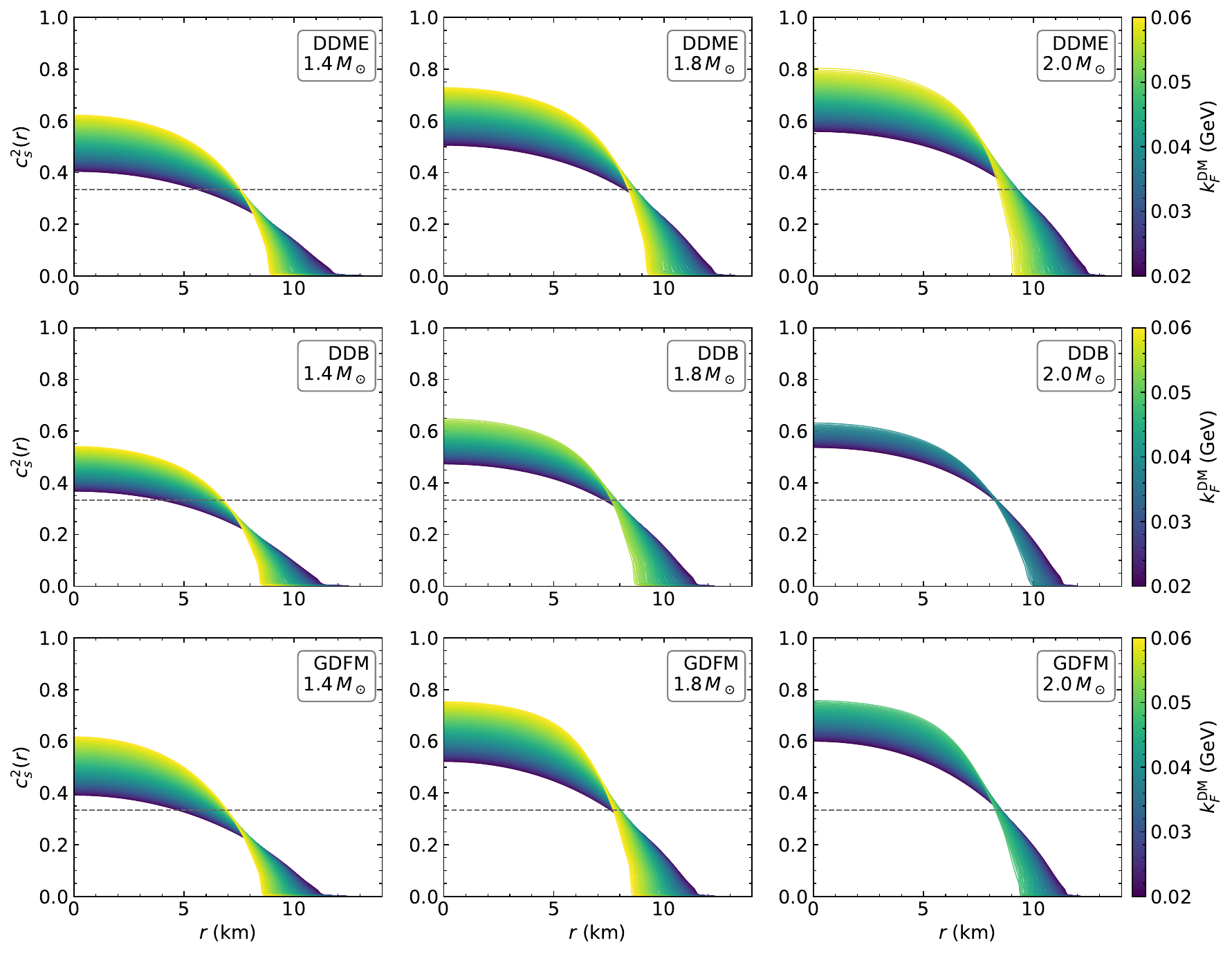}    
    \end{tabular}
    \caption{Squared sound speed $c_s^2$ as a function of the radial coordinate $r$ for NSs with masses $1.4 M_{\odot}$, $1.8 M_{\odot}$, and $2 M_{\odot}$ (left to right panels), computed within the DDME, DDB, and GDFM models (top to bottom rows). Curves are displayed for DM Fermi momenta between $0.02$ GeV and $0.06$ GeV. The horizontal dashed line marks the conformal limit $c_s^2 = \frac{1}{3}$.}
    \label{fig:3}
\end{figure*}

Fig. \ref{fig:3} presents the squared sound speed profile, $c_s^2(r)$, as a function of the radial coordinate $r$ for three stellar masses ($1.4$, $1.8$, $2.0$ $M_{\odot}$), considering the DDME, DDB, and GDFM models, and for different values of the DM Fermi momentum $k_F^{DM}$ ranging from 0.02 to 0.06
 GeV. In all three models and for all stellar masses, $c_s^2(r)$ reaches its maximum value at the stellar center ($r = 0$) and then decreases monotonically toward the surface. This trend reflects the expected behaviour that the core contains the densest, stiffest matter, while the outer layers are progressively softer. The central sound speed values are relatively large—reaching or even surpassing the conformal limit of $c_s^2 = \frac{1}{3}$, especially in the case of the DDME model, which systematically produces stiffer cores than DDB and GDFM. As $k^{DM}_F$ rises from $0.02$ to $0.06$ GeV, a star of fixed gravitational mass becomes more compact: its central density increases, so the \emph{central} value of $c_s^2$ rises systematically (for example, from $c_s^2\simeq0.41$ to $c_s^2\simeq0.62$ for the $1.4\,M_{\odot}$ DDME configuration), while the profile falls off more steeply toward the surface and terminates at a smaller radius. Consequently the $c_s^2(r)$ curves for different $k^{DM}_F$ \emph{cross}: the high-DM profiles are higher at the center but shorter in radial extent. Although the admixed DM softens the \emph{global} EoS---lowering $M_{\rm max}$, $R$ and $\Lambda$---it simultaneously drives the fixed-mass stellar core to higher density and hence to higher central stiffness; the shrinking radial region over which $c_s^2$ remains significant is the microscopic counterpart of the compactification seen in the $M$--$R$ relation. Among the three models, DDME yields the broadest and most elevated $c_s^2$ profiles, whereas DDB and GDFM give rise to more compact and suppressed profiles that are nearly indistinguishable from each other---particularly at larger DM densities---despite their different density-dependent coupling parametrizations. For more massive stars ($1.8$ and $2.0$ $M_{\odot}$), the central sound speed is typically higher than in the $1.4 M_{\odot}$ configurations at the same $k^{DM}_F$. This agrees with the physical expectation that larger masses require higher
central densities. At large $k^{DM}_F$, the DM-driven rise of these already elevated central sound speeds becomes particularly strong, showing that DM exerts a more pronounced relative effect on the internal stiffness of heavier NSs.

\begin{figure*}[!th]
    \centering
    \includegraphics[width=1\linewidth]{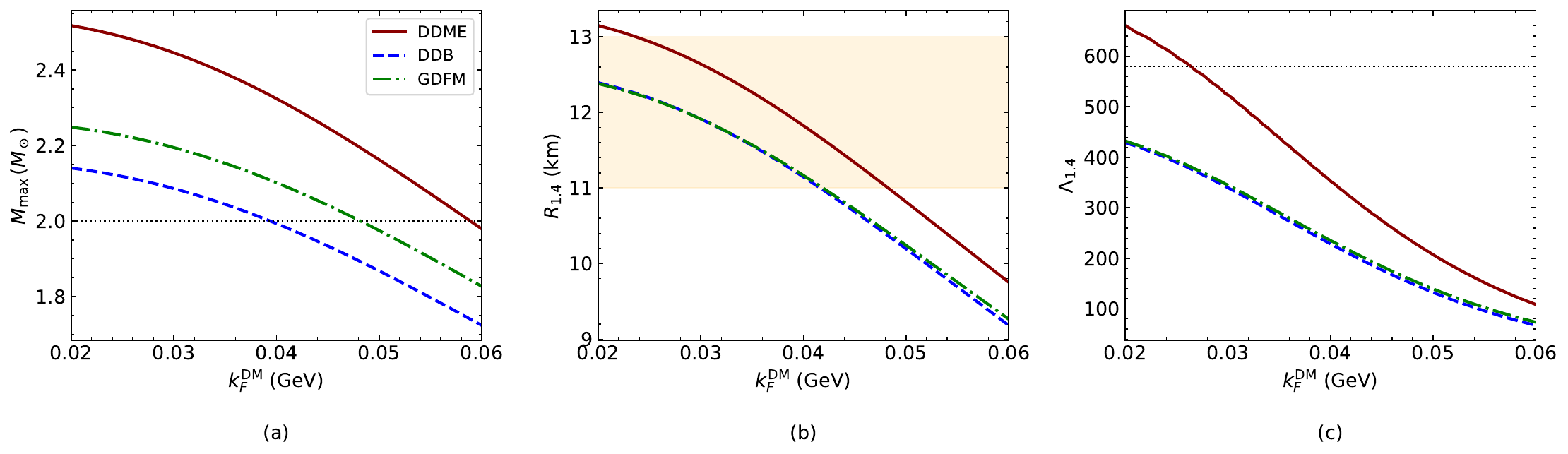}
    \caption{Macroscopic observables as functions of the DM Fermi momentum $k_F^{\rm DM}$ for the three
    models. \emph{Left}: maximum gravitational mass $M_{\rm max}$ (dotted line: the $2\,M_\odot$ pulsar
    constraint). \emph{Middle}: radius of a canonical $1.4\,M_\odot$ star $R_{1.4}$ (shaded band: the NICER
    $11$--$13$~km range). \emph{Right}: dimensionless tidal deformability $\Lambda_{1.4}$ (dotted line: the
    GW170817 bound $\Lambda_{1.4}\le580$).}
    \label{fig:4}
\end{figure*}
Figure~\ref{fig:4} collects the three macroscopic observables as functions of the DM Fermi momentum
$k_F^{\rm DM}$ for the three models. The left panel shows the maximum gravitational mass $M_{\rm max}$, which
decreases monotonically with $k_F^{\rm DM}$ in every model as the heavy, nearly pressureless DM component
softens the EoS and lowers the mass that can be supported against collapse. DDME, the stiffest functional,
yields the largest masses, falling from $2.52\,M_\odot$ at $k_F^{\rm DM}=0.02$~GeV to $1.98\,M_\odot$ at
$0.06$~GeV and meeting the two-solar-mass pulsar limit up to $k_F^{\rm DM}\simeq0.059$~GeV. The softer DDB
and GDFM models start at $2.14$ and $2.25\,M_\odot$ and drop below $2\,M_\odot$ already at
$k_F^{\rm DM}\simeq0.039$ and $0.048$~GeV, respectively, so the $2\,M_\odot$ requirement sets a direct,
model-dependent upper bound on the DM content.

The middle panel shows the radius of the canonical $1.4\,M_\odot$ star, $R_{1.4}$, which NICER constrains to
roughly $11$--$13$~km. $R_{1.4}$ falls monotonically and steeply with $k_F^{\rm DM}$ as the star is made
more compact: DDME from $13.15$ to $9.75$~km, DDB from $12.39$ to $9.18$~km, and GDFM from $12.38$ to
$9.26$~km. All three remain within the NICER band up to $k_F^{\rm DM}\simeq0.042$--$0.048$~GeV (DDME being
consistent over the widest interval, to $\simeq0.048$~GeV) and fall below the $11$~km floor beyond it,
giving a second, independent upper bound on the DM content. Already at the smallest $k_F^{\rm DM}$ the
three radii sit at or below the upper NICER edge, so no fine tuning of the DM content is needed for
radius consistency.

The right panel shows the tidal deformability of the canonical star, $\Lambda_{1.4}$, together with the
GW170817 bound $\Lambda_{1.4}\le580$. $\Lambda_{1.4}$ drops steeply with $k_F^{\rm DM}$ --- from $661$,
$429$ and $432$ at $0.02$~GeV to $109$, $67$ and $73$ at $0.06$~GeV for DDME, DDB and GDFM, respectively.
The stiff DDME functional already lies in mild tension with GW170817 in the DM-free limit
($\Lambda_{1.4}\simeq733$ at $k_F^{\rm DM}=0$) and is brought into agreement only for
$k_F^{\rm DM}\gtrsim0.026$~GeV; for this model the tidal data therefore bound the DM content from
\emph{below}, complementing the upper bounds set by $M_{\rm max}$ and $R_{1.4}$. The softer DDB and GDFM
($\Lambda_{1.4}\simeq478$--$479$ at $k_F^{\rm DM}=0$) satisfy the bound across the entire range, so for
these functionals the binding limits on the DM content come from $M_{\rm max}$ and $R_{1.4}$ alone.

A striking feature of Fig.~\ref{fig:4} is that, although the three models differ in their absolute
predictions, the \emph{relative} DM-induced changes are nearly universal. Normalising each observable to
its DM-free ($k_F^{\rm DM}=0$) value, the three models collapse onto common curves to within
$\lesssim3\%$ over the whole range, well represented by
\begin{align}
\frac{M_{\rm max}(k_F^{\rm DM})}{M_{\rm max}(0)} &\simeq 1 + 0.96\,\hat{k} - 75\,\hat{k}^2, \nonumber\\
\frac{R_{1.4}(k_F^{\rm DM})}{R_{1.4}(0)} &\simeq 1 + 0.80\,\hat{k} - 91\,\hat{k}^2,
\label{eq:scaling}
\end{align}
with $\hat{k}\equiv k_F^{\rm DM}/{\rm GeV}$, while the tidal deformability follows the compact relation
$\Lambda_{1.4}/\Lambda_{1.4}(0)\simeq[R_{1.4}/R_{1.4}(0)]^{6.1}$, reflecting the $\Lambda\propto R^5$
scaling combined with the additional DM-driven reduction of the Love number. These near-universal laws
allow the DM-induced shift of any of the three observables to be estimated directly from $k_F^{\rm DM}$,
independent of the specific density-dependent parametrization.

\begin{figure*}[!th]
    \centering
    \includegraphics[width=0.95\linewidth]{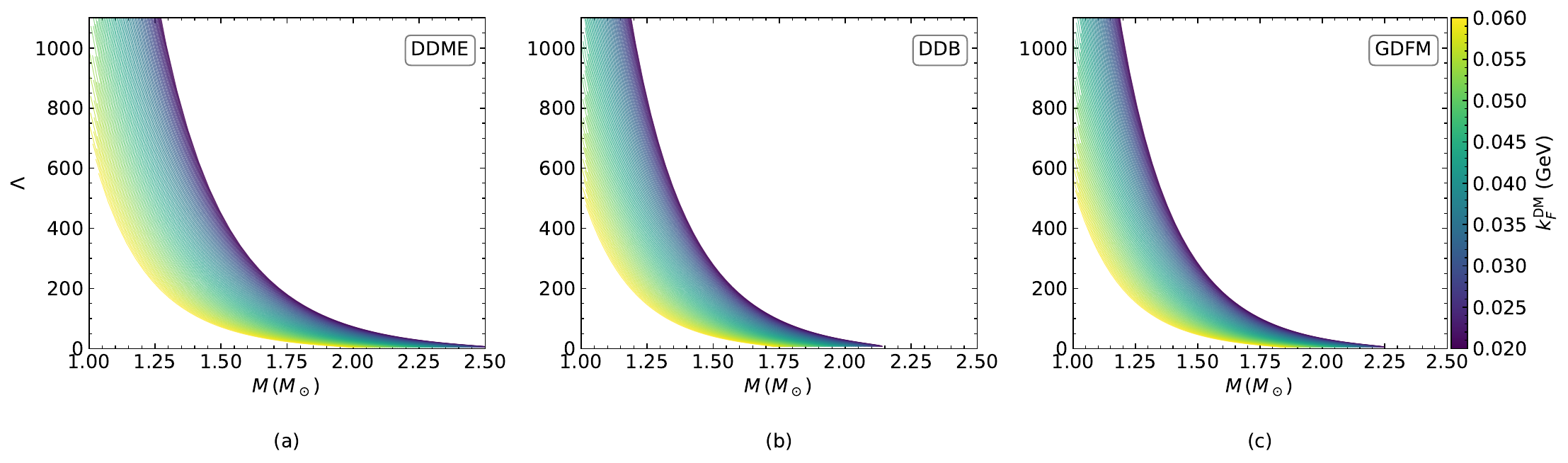}
    \caption{Dimensionless tidal deformability $\Lambda$ as a function of NS mass for the three models,
    for DM Fermi momenta $k_F^{\rm DM}=0.02$--$0.06$~GeV.}
    \label{fig:5}
\end{figure*}
Figure~\ref{fig:5} presents the full $\Lambda$--$M$ relation for the three
models over
$k_F^{\rm DM}=0.02$--$0.06$~GeV. In every model $\Lambda$ falls steeply with mass, since more massive
stars are more compact and harder to deform, and the whole family of curves shifts to lower $\Lambda$ as
$k_F^{\rm DM}$ grows --- a direct consequence of the DM-induced softening. DDME yields the largest
$\Lambda$ at fixed mass, in line with its stiffer EoS, while DDB and GDFM give lower, nearly
indistinguishable values despite their differing coupling parametrizations.
\vspace{0.8cm}
\subsection{Distinguishing the dark-matter imprint from nucleonic-EoS uncertainty}
\label{sec:dRdM}
The macroscopic effects of admixed dark matter described above --- a reduction of $M_{\max}$, $R_{1.4}$
and $\Lambda_{1.4}$ --- are qualitatively degenerate with a softening of the purely nucleonic equation of
state. A more demanding question is therefore whether the DM imprint can be separated from the present
uncertainty of the nucleonic EoS using \emph{measurable} quantities. To this end we work with two
genuine observables evaluated at fixed gravitational mass ($1.4$, $1.8$ and $2.0\,M_\odot$): the local
slope of the mass--radius relation, $dR/dM$, accessible from NICER-type mass--radius determinations, and
the dimensionless tidal deformability $\Lambda$, measured in gravitational-wave inspirals. The local
slope of the $M(R)$ curve has been proposed as a diagnostic of the underlying dense-matter composition:
the onset of a new, softening degree of freedom --- such as hyperons or deconfined quark matter ---
drives $dR/dM$ more negative and correlates with a smaller high-mass radius \citep{Ferreira:2024hxc}.
In the dark-matter context, Thakur et al.\ \citep{Thakur:2024btu} examined the $dR/dM$ derivative
at a fixed mass to distinguish a neutron-decay dark-matter scenario within a nonlinear RMF model. Our
analysis extends this feature-based approach in three respects: we treat a Higgs-portal fermionic-DM
coupling, work systematically across three density-dependent functionals, and study the imprint on both
$dR/dM$ and $\Lambda$ as a function of mass against a full Bayesian nucleonic posterior. As the
nucleonic reference we use the publicly released Bayesian DDRMF posterior of Cartaxo et al.\
(2026)\,\citep{Cartaxo:2025jpi} ($\sim\!24{,}000$ DM-free EoS per model, conditioned on nuclear, chiral-EFT,
perturbative-QCD and multimessenger data); its spread at each mass defines the nucleonic $1\sigma$ and
$2\sigma$ bands.

\begin{figure*}[!th]
    \centering
    \includegraphics[width=0.96\linewidth]{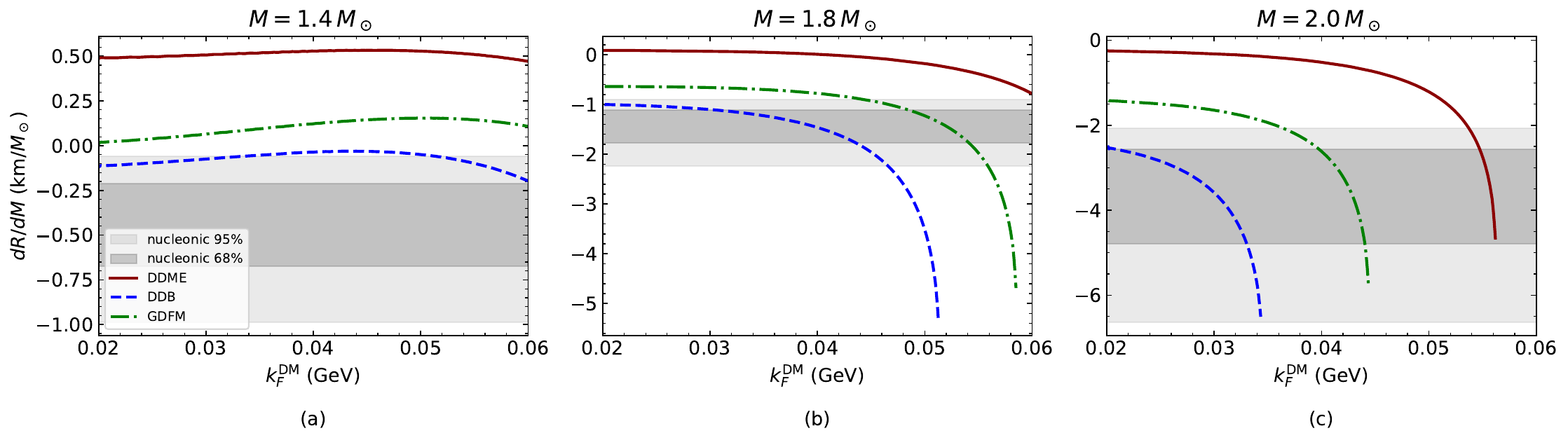}
    \caption{Slope of the mass--radius relation $dR/dM$ at fixed gravitational mass
    ($1.4$, $1.8$, $2.0\,M_\odot$) as a function of the DM Fermi momentum $k_F^{\rm DM}$ for the three
    models. Grey bands show the $1\sigma$ (dark) and $2\sigma$ (light) spread of the DM-free nucleonic
    posterior of Cartaxo et al.\ (2026)\,\citep{Cartaxo:2025jpi}.}
    \label{fig:dRdM_kF}
\end{figure*}

Figure~\ref{fig:dRdM_kF} shows $dR/dM$ versus $k_F^{\rm DM}$ together with these bands, and the response is
strikingly mass-dependent. At $1.4\,M_\odot$ the slope is almost \emph{independent} of $k_F^{\rm DM}$ in all
three models and stays within the broad nucleonic band, so the canonical-mass slope is a weak discriminator.
At $1.8\,M_\odot$ this stability is lost for the softer DDB and GDFM, whose slope bends downward with
increasing $k_F^{\rm DM}$, while the stiff DDME remains nearly flat; at $2.0\,M_\odot$ the bending steepens
further and now affects all three models. The physics is that $dR/dM$ at fixed mass measures how close that
star sits to its maximum-mass configuration, where the $M$--$R$ curve turns over and $dR/dM\to-\infty$.
Because the captured DM lowers $M_{\rm max}$, a star of fixed mass is driven toward this turning point, and
the shift grows both with the stellar mass (nearer $M_{\rm max}$) and with the softness of the functional
(lower $M_{\rm max}$): a $1.4\,M_\odot$ star stays far from the turn-over for every model and is insensitive,
whereas a $2.0\,M_\odot$ star --- already close to $M_{\rm max}$ for the softer functionals --- shows the
steepest DM-driven bending, the stiff DDME remaining comparatively flat thanks to its high $M_{\rm max}$.
Only for the heavier $1.8$ and $2.0\,M_\odot$ branches, therefore, does the DM drive $dR/dM$ cleanly out of
the nucleonic band.

\begin{figure*}[!th]
    \centering
    \includegraphics[width=0.96\linewidth]{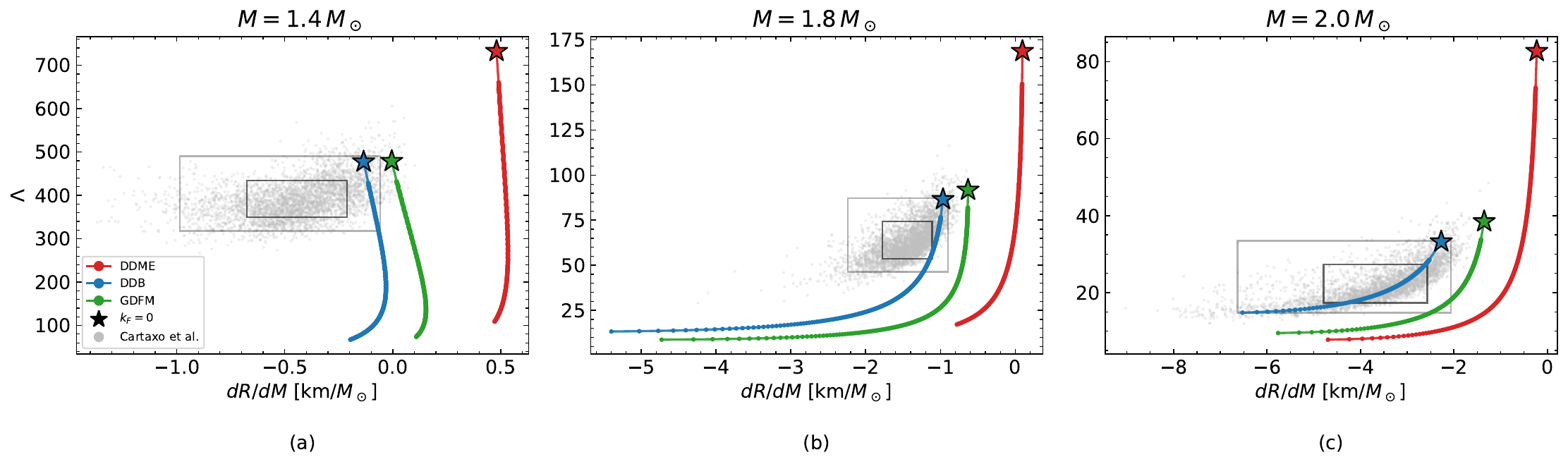}
    \caption{Dark-matter imprint in the plane of two observables, the tidal deformability $\Lambda$ versus
    the mass--radius slope $dR/dM$, at fixed mass ($1.4$, $1.8$, $2.0\,M_\odot$). Grey points and boxes:
    Cartaxo nucleonic posterior and its $1\sigma$/$2\sigma$ regions. Stars: each model's DM-free
    ($k_F^{\rm DM}=0$) value; colour-coded tracks: the imprint for $k_F^{\rm DM}=0.02$--$0.06$~GeV. The DM
    tracks leave the nucleonic region predominantly by lowering $\Lambda$ well below the nucleonic floor.}
    \label{fig:dRdM_lam}
\end{figure*}

The tidal deformability is far more discriminating, and is the central result of this analysis.
Figure~\ref{fig:dRdM_lam} shows the DM imprint in the $dR/dM$--$\Lambda$ plane, both axes observable. The
stars mark each model's DM-free ($k_F^{\rm DM}=0$) prediction and the colour-coded tracks follow the
imprint as $k_F^{\rm DM}$ increases. In every panel the tracks plunge in $\Lambda$ and exit the nucleonic
posterior cloud downward. To quantify the separation we measure, for each model and mass, the shift of the
DM-lowered $\Lambda$ from the mean of the Cartaxo posterior in units of that posterior's $1\sigma$ width
$\sigma_{\rm nuc}$. At $1.4\,M_\odot$ and $k_F^{\rm DM}=0.06$~GeV this shift reaches
$(\mathrm{mean}_{\rm nuc}-\Lambda_{\rm DM})/\sigma_{\rm nuc}\simeq7.8,\,7.5,\,7.7$ for DDME, DDB and GDFM
respectively --- a model-independent $\simeq\!8\sigma$ displacement below the band. The DM track crosses
the nucleonic $1\sigma$ boundary already at $k_F^{\rm DM}\simeq0.041,\,0.027,\,0.030$~GeV (DDME, DDB, GDFM)
and the $2\sigma$ boundary at $k_F^{\rm DM}\simeq0.043,\,0.032,\,0.034$~GeV. In observational
terms, a tidal deformability at fixed mass measured below the nucleonic floor --- e.g.\ $\Lambda_{1.4}$
appreciably smaller than the $2\sigma$ lower edge of the combined band, $\Lambda_{1.4}\sim\!320$ found here
--- could not be reproduced by any nucleonic EoS in the posterior and would signal a non-nucleonic,
softening component. We emphasise that such a measurement would identify the \emph{presence} of extra
softening beyond the nucleonic sector, but not its nature: hyperonic degrees of freedom
\citep{Huang:2024rvj}, a quark/hybrid phase transition \citep{Annala:2019puf}, or
$\Delta$-isobars \citep{Li:2019tjx} soften the high-density EoS in the same direction and would leave
qualitatively similar imprints on $M_{\rm max}$, $R_{1.4}$, $\Lambda_{1.4}$ and $c_s^2$.
Dark matter is thus one well-motivated candidate among several; discriminating it from hadronic exotica
would require complementary information (e.g.\ its distinct effect on the crust, cooling, or the
mass at which the softening sets in) and lies beyond the scope of the present study.
We stress that the relevant, parametrization-independent statement is the magnitude of this
DM-induced \emph{shift} relative to the nucleonic width, since the three representative parametrizations
are intrinsically stiff and sit at (DDB, GDFM) or above (DDME, whose DM-free $\Lambda_{1.4}\simeq733$ lies
well above the band) its upper edge even in the DM-free limit; the shift is therefore reported as a
displacement from the posterior mean rather than an absolute distance.

\begin{figure*}[!th]
    \centering
    \includegraphics[width=0.96\linewidth]{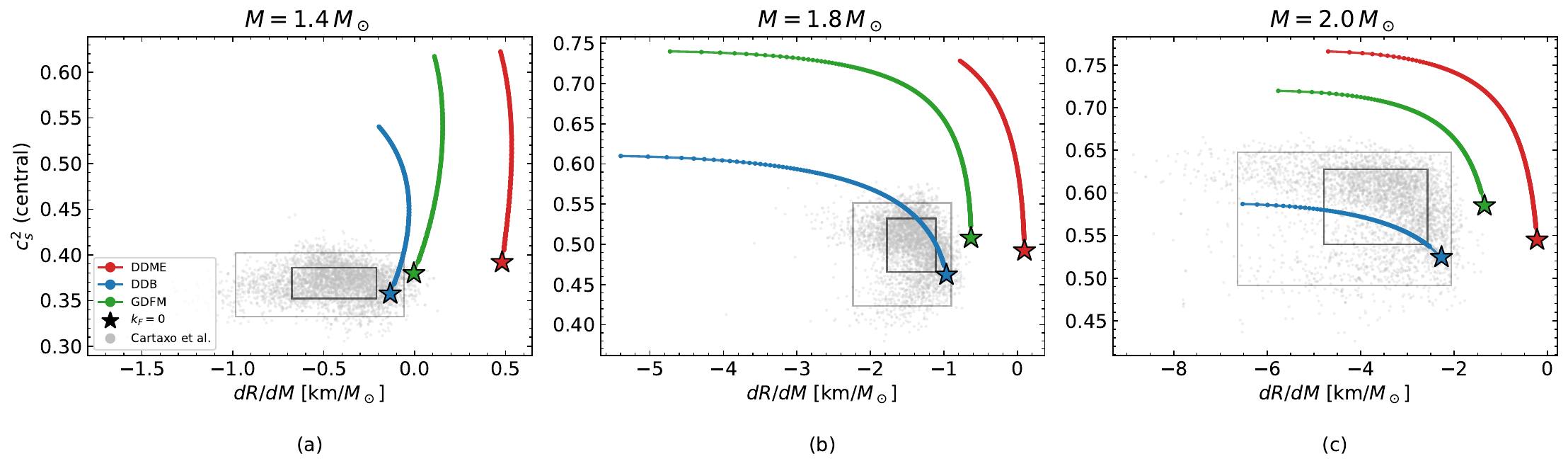}
    \caption{As Fig.~\ref{fig:dRdM_lam}, but for the central squared sound speed $c_s^2$ versus the
    mass--radius slope $dR/dM$ at fixed mass. The nucleonic $c_s^2$ band is intrinsically narrow, so the
    DM-induced rise of $c_s^2$ is the largest relative shift of any quantity considered here. The central
    $c_s^2$ is a microscopic quantity rather than a direct observable; it is shown to expose the physical
    mechanism --- a DM-driven increase of the central density beyond the nucleonic reach --- behind the
    collapse of the (observable) tidal deformability in Fig.~\ref{fig:dRdM_lam}.}
    \label{fig:dRdM_cs2}
\end{figure*}

The underlying mechanism is the same one that makes the central sound speed rise with DM (Fig.~\ref{fig:3}):
because the DM contributes energy density with negligible pressure, it drives a star of fixed
gravitational mass to a compactness --- and hence a central density --- that no nucleonic EoS in the
posterior can reach, which simultaneously raises the central $c_s^2$ and collapses the tidal deformability.
The same analysis carried out in the $dR/dM$--$c_s^2$ plane is shown in Fig.~\ref{fig:dRdM_cs2}: because
nucleonic matter (NM) at $2$--$3\,n_0$ is near-conformal, the nucleonic $c_s^2$ band at fixed mass is
intrinsically narrow, so the DM-driven rise of the central $c_s^2$ stands out even more sharply, exceeding
the nucleonic $1\sigma$ width by more than an order of magnitude already at $1.4\,M_\odot$. We emphasize,
however, that the central $c_s^2$ is a microscopic quantity that is not directly measured, whereas
$\Lambda$ is a gravitational-wave observable. The tidal deformability at fixed mass therefore emerges as
the most promising \emph{observable} signature of a Higgs-portal DM component --- with the high-mass
mass--radius slope $dR/dM$ providing complementary leverage and the central sound speed providing the
physical interpretation.

Two caveats delimit the strength of this statement. First, the significance we quote measures the DM shift
against the \emph{width of the nucleonic posterior}, not against the measurement uncertainty of a given
gravitational-wave event; the two coincide only in the (future) regime where the observational error on
$\Lambda$ at fixed mass has shrunk below $\sigma_{\rm nuc}$. For a GW170817-class measurement the tidal
error bars remain broad, so the practical discriminating power quoted here should be read as the
information available once next-generation detectors (or a stacked population) reach the requisite
precision; a dedicated Fisher/injection forecast is left to future work. Second, because the three
representative functionals are stiff and lie at or above the upper edge of the band even without DM, the
robust, parametrization-independent quantity is the \emph{differential} displacement of the DM track from
its own DM-free starting point relative to $\sigma_{\rm nuc}$, rather than the absolute distance from the
band. The two definitions agree to within a factor of order unity for the softer DDB and GDFM
functionals; DDME, whose DM-free $\Lambda_{1.4}$ already exceeds the band, is best read differentially. We also note that the Cartaxo posterior is itself conditioned in part on the GW170817
tidal measurement, so the band and the GW170817 constraint are not statistically independent; a fully
rigorous treatment would rebuild the band with GW170817 excluded, which we defer to a dedicated study.

\section{Conclusions}\label{sec:conclusions}

We have studied the structure of fermionic-dark-matter-admixed neutron stars in three density-dependent
relativistic mean-field models --- DDME, DDB and GDFM --- with the dark matter (DM) coupled to nucleons
through the Standard-Model Higgs portal. Treating the DM Fermi momentum $k_F^{\rm DM}$ as a control
parameter over $0.02$--$0.06$~GeV, we computed the equation of state, the mass--radius relation, the
maximum mass, the radial sound-speed profile, and the tidal deformability, and we
confronted the results with current multi-messenger constraints. Our main findings are as follows.

(i) In all three models the captured DM, behaving as a heavy and nearly pressureless Fermi component,
softens the equation of state (EoS): with increasing $k_F^{\rm DM}$ the maximum mass, the canonical radius
$R_{1.4}$ and the tidal deformability $\Lambda_{1.4}$ decrease monotonically, while the star becomes more
compact. DDME (the stiffest functional) is the most resilient, retaining $M_{\rm max}>2\,M_\odot$ up to
$k_F^{\rm DM}\simeq0.059$~GeV, whereas the softer DDB and GDFM fall below $2\,M_\odot$ already at
$k_F^{\rm DM}\simeq0.039$ and $0.048$~GeV, respectively.

(ii) The combination of the two-solar-mass pulsar limit and the NICER radius range
($R_{1.4}\simeq11$--$13$~km) places a model-dependent upper bound on the DM content: requiring
consistency with both, the allowed DM Fermi momentum does not exceed $k_F^{\rm DM}\simeq0.04$--$0.05$~GeV
($n_{\rm DM}\lesssim\mathrm{few}\times10^{-3}\,n_0$), the precise value depending on the nucleonic functional. The GW170817
tidal bound $\Lambda_{1.4}\le580$ is comfortably satisfied by the softer DDB and GDFM models throughout
the explored range, where it is the weakest of the three constraints. For the stiff DDME functional the
roles reverse: its DM-free tidal deformability ($\Lambda_{1.4}\simeq733$) already exceeds the bound, so the
tidal data \emph{require} a nonzero DM content ($k_F^{\rm DM}\gtrsim0.026$~GeV) and bound it from below,
while the two-solar-mass and NICER radius limits bound it from above.

(iii) DDB and GDFM yield closely similar radii, tidal responses and sound-speed profiles despite their
different coupling parametrizations, whereas DDME systematically predicts larger radii and a stiffer,
higher central sound speed, confirming that the choice of density-dependence functional, and not only the
saturation properties, controls the high-density behaviour.

(iv) Using the publicly released Cartaxo et al.\ (2026) Bayesian DDRMF posterior as the nucleonic
uncertainty band, we assessed whether the DM imprint is separable from the present uncertainty of the
nucleonic EoS using measurable quantities. The tidal deformability at fixed gravitational
mass emerges as a sharp, \emph{observable} discriminator: the DM-induced reduction of $\Lambda$ --- a
consequence of the DM driving the star to a compactness, and hence a central density, unreachable by
purely nucleonic matter --- reaches $\simeq\!8$ times the nucleonic $1\sigma$ width at $1.4\,M_\odot$ and
$k_F^{\rm DM}=0.06$~GeV (model-independently $7.5$--$7.8\sigma$) and drives $\Lambda$
below the nucleonic $1\sigma$ band for $k_F^{\rm DM}\gtrsim0.03$--$0.04$~GeV. The same mechanism raises the central sound
speed, but the latter is a microscopic, non-observable quantity, whereas $\Lambda$ is measured directly in
gravitational-wave inspirals. By contrast, the mass--radius slope $dR/dM$ at $1.4\,M_\odot$ is largely
degenerate with the nucleonic uncertainty and becomes informative only for the heavier
($1.8$--$2.0\,M_\odot$) branch. This identifies the tidal deformability at fixed mass as the most
promising observable signature of a Higgs-portal DM component, and motivates folding the DM degree of
freedom directly into Bayesian EoS inference as a natural next step. We caution, however, that this
signature diagnoses the presence of extra softening beyond the nucleonic sector rather than dark matter
specifically: hyperons \citep{Huang:2024rvj}, deconfined quark matter \citep{Annala:2019puf},
or other exotic degrees of freedom such as $\Delta$-isobars \citep{Li:2019tjx} soften the
high-density EoS in the same sense and would produce qualitatively similar shifts in $\Lambda$, $R$ and
$M_{\rm max}$. Disentangling a dark component from hadronic exotica will require additional, physically
distinct observables, which we leave to future work. In this sense the present study is best regarded as
a proof of concept: it demonstrates that a Higgs-portal dark-matter component belongs to the same class
of high-density ``softening agents'' whose imprint on the collective observables $M_{\rm max}$, $R_{1.4}$,
$\Lambda_{1.4}$ and $c_s^2$ is being explored for hadronic exotica, and it quantifies that imprint against
a Bayesian nucleonic band. We stress that this conclusion is drawn consistently across three distinct
density-dependent functionals (DDME, DDB and GDFM); the DM-induced trends and their magnitude relative to
the nucleonic band do not depend on the particular functional form of the density-dependent couplings, so
the qualitative picture is a generic property of the Higgs-portal mechanism rather than an artefact of a
single parametrization. Finally, we note that no single observable breaks the composition degeneracy on
its own: neither the $M$--$R$ curve, nor $\Lambda(M)$, nor the stellar oscillation spectrum uniquely
fingerprints dark matter. Dark-matter imprints on the non-radial oscillation modes have themselves been
studied \citep{Shirke:2023ktu,Flores:2024hts,Dey:2024vsw,Thakur:2024ejl}, and there too a
degeneracy between purely hadronic and DM-admixed configurations has been demonstrated
\citep{Shirke:2024ymc}. This reinforces the need for a multi-observable, posterior-aware strategy rather
than reliance on any single mass--radius, tidal, or oscillation-mode measurement.

\begin{acknowledgments}
M. Molla is thankful to the Inter-University Centre for Astronomy and Astrophysics (IUCAA), Pune, India, for providing research facilities during his stay there. M.Molla also sincerely acknowledges the University Grants Commission (UGC) for financial assistance through the Senior Research Fellowship (SRF). M. Kalam expresses his gratitude to ICARD, Aliah University, for the research facilities provided, as well as to the IUCAA for offering the Associateship programme. M.Kalam further thanks the Research Development Cell (RDC) at Aliah University for partial financial support under Project ID: AURDC/2024-25/PHY/07 for this research work. This work was partially supported by national funds from FCT (Fundação para a Ciência e a Tecnologia, I.P, Portugal) under project UID/04564/2025, identified by DOI 10.54499/UIDB/04564/2025, and project  2024.16290.PEX identified by DOI  identifier 10.54499/2024.16290.PEX.
\end{acknowledgments}

\section*{Data Availability Statement}
This manuscript does not include experimentally measured data; all plots are based on data generated through modeling. The generated EoSs can be obtained from the corresponding author upon reasonable request.

\vspace{5cm}

\bibliography{references}

\end{document}